\documentclass{article}
\usepackage{amsmath}
\usepackage{geometry}
\usepackage[colorlinks=true, linkcolor=blue, urlcolor=blue]{hyperref}
\usepackage{enumitem}
\usepackage{amssymb}
\usepackage{graphicx}
\usepackage{natbib}
\usepackage{amsthm}

\theoremstyle{plain}

\theoremstyle{remark}

\title{Estimation of Multivariate Functional Principal Components from Sparse Functional Data}

  \author{
    Uche Mbaka\footnote{School of Mathematics and Statistics, University College Dublin, Belfield, Dublin, Ireland.}\hspace{.3cm}and 
    Michelle Carey$^*$
    }
\date{March 2026}

\begin{document}

\maketitle

\begin{abstract}
Traditional Functional Principal Component Analysis typically focuses on densely observed univariate functional data, yet many applications, particularly in longitudinal studies, involve multivariate functional data observed sparsely and irregularly across subjects. A common approach for extracting multivariate functional principal components in such settings relies on an eigen decomposition of univariate functional principal component scores to capture cross-component correlations. We propose a novel method for the estimation of multivariate functional principal components by improving the univariate eigenanalysis through maximum likelihood estimation combined with a modified Gram–Schmidt orthonormalization. The performance of the proposed approach is evaluated against two established methods and its practical utility is demonstrated through an application to longitudinal cognitive biomarker data from an Alzheimer’s disease study and a collection of data on dairy milk yield and milk compositions from research dairy farms in Ireland.
\end{abstract}

\noindent%
\newpage
\section{Introduction} \label{sec:intro}

Functional Principal Component Analysis (FPCA) is a cornerstone of Functional Data Analysis (FDA), extending traditional Principal Component Analysis (PCA) to functional observations \citep{Ramsay_Silverman_2005, Silverman_1996, Hall_Müller_Wang_2006}. The foundational concepts were introduced by \cite{karhunen1947under} and \cite{loeve1945fonctions}, and were further developed by \cite{Dauxois_Pousse_Romain_1982}. FPCA  analyzes the covariance structure of the data to uncover the principal modes of variation, represented as eigenfunctions of the covariance operator. These modes describe the primary deviations of functions in the dataset from their mean. Scores associated with these eigenfunctions provide insight into individual variations from typical behaviour and are often used in subsequent analysis such as clustering, outlier detection, and regression  \citep{Ramsay_Dalzell_1991, Yao_Müller_Wang_2005, James_Sugar_2003}. 

While classical FPCA assumes densely and regularly observed functions on a common grid, many applications, especially in longitudinal studies, involve irregular and sparsely observed data. To address this, various methods have been developed, including those proposed by \cite{Shi_Weiss_Taylor_1996, James_2000, Rice_Wu_2001, James_Hastie_2001, James_Sugar_2003, Yao_Müller_Wang_2005, Paul_Peng_2011, Xiao_Li_Checkley_Crainiceanu_2018, Wong_Zhang_2019, Shi_Yang_Wang_Ma_Beg_Pei_Cao_2022, Mbaka_Cao_Carey_2024}. In particular, \cite{Paul_Peng_2011} introduced an approach that approximates the eigenfunctions using a finite basis of smooth functions. To ensure orthonormality of the estimated eigenfunctions, their method maximizes a restricted log-likelihood over the Stiefel manifold, which represents the set of orthonormal matrices. While this approach is theoretically appealing, it faces practical challenges. The non-convex geometry of the Stiefel manifold makes the optimization sensitive to initialization and prone to slow convergence, especially when the likelihood surface is ill-conditioned, as is often the case with sparse data. To address this, \cite{Mbaka_Cao_Carey_2024} proposed an alternative approach that enforces orthonormality utilizing a modified Gram-Schmidt procedure during the estimation process. This avoids the complexity of manifold-based optimization. Parameter estimation is then carried out by minimizing the negative log-likelihood using a standard quasi-Newton algorithm, offering a more stable optimization procedure.

Multivariate functional data occurs when each observational unit is associated with multiple functions. For example, a single patient might have functions representing their heart rate, blood pressure, and glucose levels measured over time. In this case, each patient is the observational unit, and the heart rate, blood pressure, and glucose functions are the multivariate functional data.
For each function \(k \in 1, \ldots, p\), each subject \(i\in 1,\ldots, n\) has observations \(Y^{(k)}_{ij}\) at observation points \(t^{(k)}_{ij} \in T^{(k)}\) with \(j \in 1, \ldots , m_i\). We assume that the $m_i$ observations for each function, $\{Y^{(k)}_{i,1},\ldots, Y^{(k)}_{i,m_i}\}$, are noisy realizations of a smooth curve, $X_{i}^{(k)}$, defined on $T$ and evaluated at a finite grid $\{t^{(k)}_{i,1},\ldots,t^{(k)}_{i,m_i}\}$, that is, $Y^{k}_{i,j}=X_{i}^{(k)}(t^{(k)}_{ij})+\epsilon^{(k)}_{i,j},$ where $\epsilon^{(k)}_{i,j}$ are independent and identically distributed measurement errors with mean zero and variance \(\sigma_k^2\).

One could apply univariate FPCA independently to each of the \(p\) functions, but this approach fails to account for potential correlations among the principal component scores across the functions. If the \(p\) variables are recorded on the same dense finite grid, \(\{t^{(k)}_{i,1},\ldots,t^{(k)}_{i,m_i}\}\), and the measurements $\{Y^{(k)}_{i,j}\}_{k=1}^{p}$ are in similar units, \cite{Ramsay_Silverman_2005} outlines a method for Multivariate Functional Principal Component Analysis (MFPCA) by amalgamating the \(p\) functions and then analyzing them using FPCA on the combined dataset. For functions that exhibit different scales or variability, normalization is required to address inherent heteroscedasticity. Two approaches to normalized MFPCA have been developed. The first approach, described by \cite{Jacques_Preda_2014}, begins with pointwise scaling of the \(p\) functions; FPCA is then performed on the resultant weighted, combined functions. The second approach, as suggested by \cite{Chiou_Yang_Chen_2014}, involves weighting the covariance functions of the \(p\) variables; FPCA is subsequently applied to these weighted covariance functions.

\cite{Happ_Greven_2018} introduced an MFPCA framework for dense and sparse functional data that accommodates data across different time intervals \(T^{(k)}\), for \(k=1,\ldots,p\), and dimensional domains, such as \(T^{(1)} \in \mathbb{R}\) and \(T^{(2)} \in \mathbb{R}^2\). This methodology involves performing either univariate FPCA (dense case) or Principal Analysis by Conditional Expectation PACE (sparse case) on each function \(X^{(k)}\) for \(k=1, \ldots, p\). The functional principal component scores obtained are then combined, and an eigendecomposition is applied to their covariance matrix. The multivariate principal components are subsequently derived from a weighted combination of the resulting eigenvectors of the amalgamated scores' covariance and the original univariate principal components. 

Alternatively, \cite{Li_Xiao_Luo_2020} developed a method called \texttt{mFACEs}, which extends the tensor-product B-spline covariance estimation technique originally devised in \cite{Xiao_Li_Checkley_Crainiceanu_2018} for estimating multivariate covariance functions. After estimating the covariance, they applied FPCA to the multivariate covariance function to extract the corresponding multivariate functional principal components.

We propose extending the univariate FPCA approach detailed in \cite{Mbaka_Cao_Carey_2024} to sparse multivariate functional data using the methodology suggested by \cite{Happ_Greven_2018}, aiming to enhance the accuracy of the resulting multivariate functional principal components. 

The structure of the article is as follows: Section \ref{sec:method} introduces MFPCA for sparse functional data, which can be defined across varying time intervals. It discusses the relationship between univariate and multivariate functional principal components, as described by \cite{Happ_Greven_2018}. It reviews the univariate FPCA method presented in \cite{Mbaka_Cao_Carey_2024}, and outlines our proposed extension to MFPCA. Section 3 benchmarks our proposed approach against \texttt{mFACEs}, as outlined by \cite{Li_Xiao_Luo_2020}, and the sparse \texttt{MFPCA} method detailed by \cite{Happ_Greven_2018}, using six simulated datasets. Section 4 demonstrates the application of our method to two datasets: one from Alzheimer's disease research and another based on milk yield and milk composition data obtained from farms in Ireland. The article concludes with final remarks in Section 5.

\section{MFPCA for Sparse Functional Data}\label{sec:method}

\subsection{Multivariate FPCs and their Basic Properties}

Consider each subject \(i = 1, \ldots, n\), with functional observations comprising of \(p \ \geq \  2\) functions \(X_i^{(1)}, \ldots, X_i^{(p)}\). These functions are not necessarily defined over the same domains \(\mathcal{T}^{(1)}, \ldots, \mathcal{T}^{(p)}\) and may have different argument values \(t^{(1)},\ldots,t^{(p)}\). Each function $X_i^{(k)}$ belongs to an $L^2(\mathcal{T}^{(k)})$ Hilbert space of square-integrable functions. The different functions are combined in a vector \[\mathbf{X}_{i}(\textbf{t}) = (X_{i}^{(1)}(t^{(1)}), \ldots, X_{i}^{(p)}(t^{(p)}))^T \in \mathbb{R}^p \text{ in } \mathcal{H} = \mathcal{H}^{(1)} \times \ldots \times \mathcal{H}^{(p)}\], where $\mathcal{H}^{(k)}$ is a Hilbert space of $L^2(\mathcal{T}^{(k)})$ and $\textbf{t}=(t^{(1)},\ldots,t^{(p)}) \in \mathcal{T}^{(1)} \times \ldots \times \mathcal{T}^{(p)}.$ Assume that $X_{i}(\cdot)$s are independent and identically distributed samples from a smooth process $\textbf{X}(\cdot)$. Denote the $p$-dimensional smooth mean of $\textbf{X}(\cdot)$ and covariance $\textbf{C}(\textbf{s},\textbf{t})$ elements by,
\begin{eqnarray*}
\nonumber
\mathbf{\mu}(\mathbf{t}) &=& E[\mathbf{X}(\mathbf{t})] = (E[X^{(1)}(t^{(1)})], \ldots, E[X^{(p)}(t^{(p)})]) \qquad  \forall \  \mathbf{t} \in \mathcal{T}\\
C_{kk'}(s^{(k)},t^{(k')}) &=& \textrm{Cov}[X^{(k)}(s^{(k)}),X^{(k')}(t^{(k')})] \quad \forall \ s^{(k)} \in \mathcal{T}^{(k)}, \ t^{(k')} \in \mathcal{T}^{(k')}
\end{eqnarray*}
The space \(\mathcal{H}\) is endowed with the inner product \(\langle \mathbf{f}, \mathbf{g} \rangle_{\mathcal{H}} = \sum_{k=1}^{p}\int f^{(k)}(t^{(k)}) \\g^{(k)}(t^{(k)}) dt^{(k)}\), and the norm \(||\cdot||_{\mathcal{H}} = \langle \cdot, \cdot \rangle^{\frac{1}{2}}_{\mathcal{H}}\), where \(\mathbf{f} = (f^{(1)}, \ldots, f^{(p)})\) and \(\mathbf{g} = (g^{(1)}, \ldots, g^{(p)})\) are elements of \(\mathcal{H}\). 

The covariance operator $\Gamma:\mathcal{H} \rightarrow \mathcal{H}$ with associated kernel $C(\mathbf{s}, \mathbf{t})$ is defined such that for any $\mathbf{f}\in \mathcal{H}$, the $k$th element of $\Gamma f$ is given by 
\begin{equation*}\label{eqn:cov_op}
    (\Gamma f)^{(k')}(t^{(k')}) := \sum^p_{k = 1}\int_{\mathcal{T}^{(k)}} C_{kk'}(s^{(k)},t^{(k')}) f^{(k)}(s^{(k)})ds^{(k)} = \left<C_{.k'}(.,t^{(k')}),\mathbf{f}\right>_{\mathcal{H}},
\end{equation*}
for $t^{(k)'} \in \mathcal{T}^{(k)'}$. The operator \(\Gamma\) is assumed to be a linear, self-adjoint, compact, and positive integral operator on \(\mathcal{H}\). 

By the Hilbert-Schimdt theorem \citep{reed1980methods}, there exists a complete orthornormal basis of eigenfunctions $\psi_l \in \mathcal{H}, l \in \mathbb{N}$ of $\Gamma$ such that 
\begin{equation}\label{eqn:eigen_eqn}
        (\Gamma \psi_l)^{(k)}(t^{(k)}) := \sum^p_{k = 1}\int_{\mathcal{T}^{(k)}} C_{kk'}(s^{(k)},t^{(k')}) \psi_l^{(k)}(s^{(k)})ds^{(k)} = \eta_l \psi_l^{(k)}(t^{(k)}),
\end{equation}
where $\eta_l$ is the \(l\)-th eigenvalues for the multivariate eigenvector $\boldsymbol{\psi}_l=(\psi_l^{(1)},\\\cdots,\psi_l^{(p)})^\top$ and thus $\eta_1 \geq \eta_2 \geq \ldots \geq 0$. Then the multivariate version of the Mercer's theorem \citep{mercer1909xvi} is
\begin{equation*}\label{eqn:Mercer}
    C_{kk'}(s^{(k)},t^{(k')}) = \sum^{\infty}_{l = 1} \eta_l \psi^{(k)}_l(s^{(k)}) \psi^{(k')}_l(t^{(k')}). 
\end{equation*}
The function $\mathbf{X(t)}$ has a multivariate Karhunen–Loève representation given as 
\begin{equation*}\label{eqn:KL}
    \mathbf{X(t)} = \mu(\mathbf{t}) + \sum^{\infty}_{l = 1}\rho_l \boldsymbol{\psi}_l(\mathbf{t}), \quad \textbf{t} \in \mathcal{T},
\end{equation*}
where the random variable $\rho_l = \langle \textbf{X}-\boldsymbol{\mu},\boldsymbol{\psi}_l \rangle_\mathcal{H}$ are the multivariate principal component scores with the properties $E(\rho_l) = 0, \ Var(\rho_l) = E(\rho_l^2) = \eta_l, \ Cov(\rho_l, \rho_{l'}) = 0$, if $l\neq l'$ \citep{saporta1981methodes}. 

The multivariate Karhunen-Lo\'eve representation has an interpretation analogous to the univariate case. The \(l\)-th eigenvalue, \(\eta_l\), quantifies the amount of variability in the function \(\textbf{X}(\mathbf{t})\) explained by the \(l\)-th multivariate functional principal component, \(\boldsymbol{\psi}_l(\mathbf{t})\). The multivariate functional principal component scores \(\{\rho_l\}_{l=1}^{\infty}\) act as weights for \(\{\boldsymbol{\psi}_l(\mathbf{t})\}_{l=1}^{\infty}\) in the Karhunen-Lo\'eve expansion of \(\textbf{X}(\mathbf{t})\). As these eigenvalues \(\{\eta_l\}_{l=1}^{\infty}\) diminish towards zero, the corresponding leading eigenfunctions increasingly capture the most significant features of \(\textbf{X}(\mathbf{t})\).

 In practice, \(\mathbf{X}\) is approximated by truncating the Karhunen-Lo\'eve expansion at \(M\). That is
\begin{equation}\label{eqn:trunKL}
     \mathbf{X}_{\lceil M \rceil} (\mathbf{t}) :=  \mu(\mathbf{t}) + \sum^M_{l=1} \rho_l \boldsymbol{\psi}_l(\mathbf{t}), \quad \mathbf{t} \in \mathcal{T}.
\end{equation}
Identifying an appropriate truncation lag \( M \) is a well-known problem in FPCA. A widely accepted approach is to select the smallest \( M \) that cumulatively explains a pre-specified percentage of the total variance in the dataset, typically 90\%. For further details, see \cite{Ramsay_Silverman_2005}.

When the functions exhibit significant differences in ranges or amounts of variation, weights are necessary to derive multivariate functional principal components that offer meaningful interpretations \citep{Chiou_Yang_Chen_2014}. The weighted version of the inner product $\langle \mathbf{f}, \mathbf{g} \rangle_{\mathcal{H}}$ and the covariance operator in (\ref{eqn:eigen_eqn}) is provided in Appendix \ref{Weighted}.

 \subsection{The relationship between Univariate and Multivariate FPCA}\label{Happ}
 
 There exists a relationship between the univariate and the multivariate FPCA for finite Karhunen–Loève expansions as detailed in \cite{Happ_Greven_2018}. Denote the univariate Karhunen–Loève expansion of each element \(X^{(k)}\) of \(\mathbf{X}\), based on observations \(\{\{t^{(k)}_{i,j},Y^{(k)}_{i,j}\}_{j=1}^{m_i}\}_{i=1}^{n}\), as:
 \begin{equation}\label{U_Expan}
     X_{i}^{(k)}(t^{(k)}) = \mu^{(k)}(t^{(k)}) + \sum^{M_k}_{q=1} \xi^{(k)}_{i,q} \phi^{(k)}_q(t^{(k)}),
 \end{equation}
where in the context of the \(k\)-th function, \(\mu^{(k)}(t^{(k)}) = E[X^{(k)}(t^{(k)})]\) represents the mean across all individuals, \(\{\phi^{(k)}_q(t^{(k)})\}_{q=1}^{M_k}\) are the univariate eigenfunctions evaluated at the argument values \(t^{(k)}\), and \(\{\{\xi^{(k)}_{i,q}\}_{q=1}^{M_k}\}_{i=1}^{n}\) are the univariate principal component scores. \(M_k\) denotes the truncation point for truncating the Karhunen–Loève expansion of the \(k\)-th function \(X^{(k)}\), with \(M \leq \sum_{k=1}^p M_k =: M_+\).

Define the matrix \(\mathbf{Z} \in \mathbb{R}^{M_+\times M_+}\) as the covariance matrix of the combined scores for each function \(k=1,\ldots,p\), where each row contains the terms \(\textrm{Cov}(\xi_{i,1}^{(1)}, \ldots, \xi_{i,M_1}^{(1)}, \ldots, \xi_{i,1}^{(p)},\ldots,\xi_{M_p}^{(p)})\).
Perform a matrix eigenanalysis on $\mathbf{Z}$ to obtain the eigenvalues $\eta_{l}$ and orthonormal eigenvectors $\textbf{c}_l$. 
The multivariate eigenfunctions in (\ref{eqn:trunKL}) can be written as
\begin{equation}\label{eqn:MFeigFun}
    \psi_l^{(k)}(t^{(k)}) = \sum^{M_k}_{j = 1} [\textbf{c}_l]_j^{(k)} \phi_j^{(k)}(t^{(k)}), \quad t^{(k)} \in \mathcal{T}^{(k)}, \quad l = 1, \ldots, M,
\end{equation}
where $[\textbf{c}_l]^{(k)} \in \mathbb{R}^{M_k}$ denotes the $k$-th block of the (orthonormal) eigenvector $\textbf{c}_l$ associated with the eigenvalue $\eta_l$. Similarly, the corresponding scores in (\ref{eqn:trunKL}) can be written as
\begin{equation}\label{eqn:MFscores}
    \rho_{i,l} = \sum^p_{k=1} \sum^{M_k}_{j = 1} [\textbf{c}_l]_j^{(k)} \xi_{i,j}^{(k)}.
\end{equation}
Thus, the Multivariate Functional Principal Components can be estimated from linear combinations of the univariate Functional Principal Components. The asymptotic properties of the estimators in (\ref{eqn:MFeigFun}) and (\ref{eqn:MFscores}) are detailed in \cite{Happ_Greven_2018}.

\subsection{Univariate Functional Principal Component Analysis for Sparse Functional Data}\label{FPCASp}

We employ the method proposed by \cite{Mbaka_Cao_Carey_2024}, outlined briefly in this section, to estimate univariate functional principal components from sparse functional data \(\{t^{(k)}_{i,j}, Y^{(k)}_{i,j}\}_{k=1}^{p}\) across \(j=1,\ldots,m_i\) and \(i=1,\ldots,n\). The observed datum \(Y^{(k)}_{ij}\), representing the \(i\)-th subject's measurement at the \(j\)-th time point for the \(k\)-th response variable, is modeled as \(Y^{(k)}_{ij} = X^{(k)}_i(t^{(k)}_{ij}) + e^{(k)}_{ij}\), where \(X^{(k)}_i(t^{(k)}_{ij})\) is defined in (\ref{U_Expan}) and \(e^{(k)}_{ij}\) are independent and identically distributed measurement errors with mean zero and variance \(\sigma_k^2\). To estimate the orthonormal eigenfunctions \(\{\phi^{(k)}_q(t^{(k)})\}_{q=1}^{M_k}\) for the \(k\)-th function, we utilize the modified Gram–Schmidt (MGS) orthonormalization process. Let $\boldsymbol{\tau}^{(k)} = \{\tau^{(k)}_1,\ldots,\tau^{(k)}_{H^{(k)}}\} \subset \mathcal{T}^{(k)}$ denote a dense grid of $H^{(k)}$ quadrature points for the \(k\)-th function, with associated trapezoidal quadrature weights $\{w^{(k)}_j\}_{j=1}^{H^{(k)}}$. This defines a discrete weighted inner product
\[
\langle f, g \rangle_{w^{(k)}} = \sum_{j=1}^{H^{(k)}} w^{(k)}_j\, f(\tau^{(k)}_j)\,g(\tau^{(k)}_j),
\]
which approximates the continuous $L^2(\mathcal{T}^{(k)})$ inner product with error of order $\mathcal{O}((H^{(k)}-1)^{-2})$ for sufficiently smooth functions. The eigenfunctions are approximated by orthonormalizing a spline basis expansion. The unconstrained basis function expansion for the $q$-th eigenfunction is defined as
\[
u^{(k)}_q(\boldsymbol{\tau}^{(k)}) = \sum_{u = 1}^{U^{(k)}} \beta^{(k)}_{qu} B_{u}(\boldsymbol{\tau}^{(k)}) = \mathbf{B}^{(k)} \boldsymbol{\beta}^{(k)}_q,
\]
where \(B_{u}(\boldsymbol{\tau}^{(k)})\) is the \(u\)-th basis function evaluated on the grid $\boldsymbol{\tau}^{(k)}$, and \(\beta^{(k)}_{qu}\) is the corresponding coefficient. The basis functions $\{B_{u}(\boldsymbol{\tau}^{(k)})\}_{u=1}^{U^{(k)}}$ are typically chosen to be either B-splines or Fourier basis functions, depending on whether cyclical behaviour is expected. See \cite{Ramsay_Silverman_2005} for further details on the selection of basis functions.

A weighted MGS orthonormalization of $\{u^{(k)}_1,\ldots,u^{(k)}_{M_{k}}\}$ with respect to $\langle \cdot, \cdot \rangle_{w^{(k)}}$ produces an orthonormal set $\{\phi^{(k)}_1,\ldots,\phi^{(k)}_{M_{k}}\}$, denoted $\boldsymbol{\Phi}^{(k)} = \mathcal{M}_{W^{(k)}}(\mathbf{B}^{(k)}\boldsymbol{\beta}^{(k)})$, given by the recursion formula:
\begin{equation}\label{GramSch}
    \phi^{(k)}_{q}(\boldsymbol{\tau}^{(k)}) = \frac{u^{(k)}_q(\boldsymbol{\tau}^{(k)}) - \sum_{v=1}^{q-1} \langle u^{(k)}_q, \phi^{(k)}_v \rangle_{w^{(k)}} \phi^{(k)}_v(\boldsymbol{\tau}^{(k)})}{\left\| u^{(k)}_q(\boldsymbol{\tau}^{(k)}) - \sum_{v=1}^{q-1} \langle u^{(k)}_q, \phi^{(k)}_v \rangle_{w^{(k)}} \phi^{(k)}_v(\boldsymbol{\tau}^{(k)}) \right\|_{w^{(k)}}},
\end{equation}
where $\|\cdot\|_{w^{(k)}} = \sqrt{\langle \cdot,\cdot \rangle_{w^{(k)}}}$. Let the matrix $\boldsymbol{\Phi}^{(k)}$ with dimensions $H^{(k)} \times M_{k}$ comprise the orthonormal eigenfunctions \(\phi^{(k)}_q\), evaluated at $\boldsymbol{\tau}^{(k)},$ for $q=1, \ldots, M_{k}$, satisfying $(\boldsymbol{\Phi}^{(k)})^\top \mathbf{W}^{(k)} \boldsymbol{\Phi}^{(k)} = \mathbf{I}_{M_k}$, where $\mathbf{W}^{(k)} = \mathrm{diag}(w^{(k)}_1,\ldots,w^{(k)}_{H^{(k)}})$. The $H^{(k)} \times H^{(k)}$ reduced-rank covariance matrix evaluated on the grid is:
\begin{equation}\label{Cov_Est}
        \boldsymbol{C}^{(k)} = \boldsymbol{\Phi}^{(k)}\, \boldsymbol{\Lambda}^{(k)}\, \boldsymbol{\Phi}^{\top(k)} + \sigma^{2}_{k}\mathbf{I},
\end{equation}
where \(\boldsymbol{\Lambda}^{(k)}=\textrm{diag}(\lambda_1^{(k)},\ldots,\lambda_{M_{k}}^{(k)})\) with $\lambda_q^{(k)}$ for $q=1,\ldots,M_{k}$ denoting the corresponding eigenvalues. In order to guarantee positive estimates for the eigenvalues \(\lambda_q^{(k)}\) and the error variance \(\sigma^{2}_{k}\), we reparameterize by defining \(\gamma^{(k)} = \log \sigma_k^2\) and \(\eta_q^{(k)} = \log \lambda_q^{(k)}\).

Let $\boldsymbol{\beta}^{(k)} = [\boldsymbol{\beta}^{(k)}_{1}, \cdots , \boldsymbol{\beta}^{(k)}_{M_{k}}] \in \mathbb{R}^{U^{(k)} \times M_k}$, where $\boldsymbol{\beta}^{(k)}_{q}$ is the length-$U^{(k)}$ vector of basis coefficients for the $q$-th eigenvector of the \(k\)-th function $X^{(k)}$. To estimate $\boldsymbol{\beta}^{(k)}$, the error variance parameter $\gamma^{(k)}$, and the eigenvalue parameters $\boldsymbol{\eta}^{(k)}=\{\eta^{(k)}_q\}_{q=1}^{M_{k}}$, we minimize the average negative log-likelihood of the data conditional on \( \{t^{(k)}_{i,j}, Y^{(k)}_{i,j}\}\):
\begin{equation}\label{eqn:LLK}
    \mathcal{L}^{(k)} = -\log L(\boldsymbol{\beta}^{(k)}, \boldsymbol{\eta}^{(k)}, \gamma^{(k)}) = \frac{1}{n} \sum_{i = 1}^n \mathrm{tr} \bigl[ (\boldsymbol{C}^{(k)}_{i})^{-1} \tilde{\mathbf{y}}^{(k)}_i (\tilde{\mathbf{y}}^{(k)}_i)^\top \bigr] + \frac{1}{n} \sum_{i=1}^n \log|\boldsymbol{C}^{(k)}_{i}|,
\end{equation}
where \(\tilde{\mathbf{y}}^{(k)}_i = \mathbf{Y}^{(k)}_i - \boldsymbol{\mu}^{(k)}_i\) and $\boldsymbol{C}^{(k)}_{i}$ is obtained by evaluating the reduced-rank covariance matrix in (\ref{Cov_Est}) at the observation locations $\mathbf{t}^{(k)}_i$ for the $i$-th individual. Minimization of (\ref{eqn:LLK}) with respect to \(\{\boldsymbol{\beta}^{(k)}, \boldsymbol{\eta}^{(k)},\gamma^{(k)}\}\) is carried out using a BFGS Quasi-Newton method with a cubic line search. The required score functions admit closed-form expressions and are provided in Appendix A.2 of the supplementary material of \cite{Mbaka_Cao_Carey_2024}.

Once the eigenfunctions have been estimated on the quadrature grid, a continuous functional representation is obtained using the symmetric inverse square root $(\mathbf{G}^{(k)}_B)^{-1/2}$ of the Gram matrix $\mathbf{G}^{(k)}_B = \int_{\mathcal{T}^{(k)}} \mathbf{B}^{(k)}(t)(\mathbf{B}^{(k)}(t))^\top dt$, yielding an orthonormal basis $\tilde{\mathbf{B}}^{(k)}(t) = (\mathbf{G}^{(k)}_B)^{-1/2}\mathbf{B}^{(k)}(t)$. The orthonormal coefficient matrix $\tilde{\boldsymbol{\beta}}^{(k)}$ is then computed as $\tilde{\boldsymbol{\beta}}^{(k)} \approx (\tilde{\mathbf{B}}^{(k)})^\top \mathbf{W}^{(k)} \boldsymbol{\Phi}^{(k)}$, so that for any $t \in \mathcal{T}^{(k)}$:
\begin{equation}\label{eqn:phi_t_k}
    \boldsymbol{\Phi}^{(k)}(t) = (\tilde{\mathbf{B}}^{(k)}(t))^\top \tilde{\boldsymbol{\beta}}^{(k)}.
\end{equation}

Once eigenfunctions and eigenvalues are estimated, subject-specific principal component scores $\hat{\boldsymbol{\xi}}^{(k)}_{i}$ are computed via conditional expectation, as in \cite{Yao_Müller_Wang_2005}:
\begin{equation}\label{eqn:score_k}
    \hat{\boldsymbol{\xi}}^{(k)}_{i} = \hat{\boldsymbol{\Lambda}}^{(k)} (\hat{\boldsymbol{\Phi}}^{(k)}_{i})^\top (\hat{\boldsymbol{C}}^{(k)}_{i})^{-1} \tilde{\mathbf{y}}^{(k)}_i,
\end{equation}
where $\hat{\boldsymbol{\Phi}}^{(k)}_{i} = [\hat{\phi}^{(k)}_1(\mathbf{t}^{(k)}_i),\ldots,\hat{\phi}^{(k)}_{M_k}(\mathbf{t}^{(k)}_i)]$ is the $m_i \times M_k$ matrix of estimated eigenfunctions evaluated at the $i$-th individual's observation times, obtained via (\ref{eqn:phi_t_k}). To select the number of basis functions \(U^{(k)}\) and the number of principal components \(M_{k}\), we use the AIC:
\begin{equation}\label{eqn:GCV}
    \mathrm{AIC} = n\mathcal{L}^{(k)} + U^{(k)}M^2_k + M_k + 1.
\end{equation}
The pair \(\left(U^{(k)},M_{k}\right)\) that minimizes the AIC in (\ref{eqn:GCV}) is selected; see \cite{Mbaka_Cao_Carey_2024} for further details.

\subsection{The Multivariate FPCA for Sparse Data}

To perform MFPCA for sparse functional data, we first apply a univariate functional principal component analysis on each of the \(p\) functions, \(X_i^{(1)}, \ldots, X_i^{(p)}\), as described in Section \ref{FPCASp}. This yields a total of \(p \times M_{+}\) univariate eigenfunctions, \[\{\phi^{(1)}_1(\boldsymbol{\tau}^{(1)}),\ldots,\phi^{(1)}_{M_1}(\boldsymbol{\tau}^{(1)}),\ldots,\phi^{(p)}_1(\boldsymbol{\tau}^{(p)}),\ldots,\phi^{(p)}_{M_p}(\boldsymbol{\tau}^{(p)}) \},\] each defined as per (\ref{GramSch}). Additionally, it produces \(p \times M_{+}\) associated eigenvalues \[\{\lambda^{(1)}_{1},\ldots,\lambda^{(1)}_{M_1},\ldots,\lambda^{(p)}_{1},\ldots,\lambda^{(p)}_{M_p} \}\], computed by exponentiating the estimates of \(\{\eta_{1}^{(k)},\ldots,\eta_{M_{k}}^{(k)}\}\) for \(k=1,\ldots,p\), which are obtained by minimizing the negative log-likelihood in (\ref{eqn:LLK}). The number of univariate principal components for each function $M_{1},\ldots,M_{p}$ is determined by minimizing the approximate AIC given in (\ref{eqn:GCV}). Subsequently, calculate the \(p\) estimated covariance functions by substituting the corresponding estimated univariate eigenfunctions and univariate eigenvalues into (\ref{Cov_Est}). The principal component scores are then derived by substituting the estimated eigenvalues, eigenvectors, and covariance function into (\ref{eqn:score_k}).

Define the matrix \(\mathbf{Z} \in \mathbb{R}^{M_+\times M_+}\) as the covariance matrix of the combined principal component scores for each function \(k=1,\ldots,p\), with each row containing the terms \(\textrm{Cov}(\xi_{i,1}^{(1)}, \ldots, \xi_{i,M_1}^{(1)}, \ldots, \xi_{i,1}^{(p)},\ldots,\xi_{i,M_p}^{(p)})\).
Conduct a matrix eigenanalysis on \(\mathbf{Z}\) to derive the eigenvalues \(\eta_{l}\) and orthonormal eigenvectors \(\mathbf{c}_l\) for $l=1,\ldots,M$. The number of multivariate principal components, denoted as \(M\), is determined based on the elbow rule as implemented by \cite{ahasverus}. The multivariate eigenfunctions are then computed by incorporating the estimated univariate eigenfunctions and \(\{\hat{\mathbf{c}}_{1},\ldots,\hat{\mathbf{c}}_{M}\}\) into (\ref{eqn:MFeigFun}). Obtain the multivariate principal component scores by substituting the estimated univariate scores and \(\{\hat{\mathbf{c}}_{1},\ldots,\hat{\mathbf{c}}_{M}\}\) into (\ref{eqn:MFscores}). The multivariate covariance is estimated by substituting the estimated multivariate eigenfunctions and eigenvalues into 
\begin{equation*}
    C_{kk'}(s^{(k)},t^{(k')}) = \sum^{M}_{l = 1} \hat{\eta}_l \hat{\psi}^{(k)}_l(s^{(k)}) \hat{\psi}^{(k')}_l(t^{(k')}). 
\end{equation*}
Finally, the estimated trajectories of the \(p\) functions over each domain \(\mathcal{T}^{(k)}\) are recovered using the truncated Karhunen–Loève expansion by substituting the estimated eigenfunctions and principal component scores into (\ref{eqn:MFeigFun}). 

\section{Simulations}
To assess the performance of our proposed approach for multivariate sparse functional principal component analysis, we compare its finite sample performance with two established methods: \texttt{mFACEs}, as described by \cite{Li_Xiao_Luo_2020}, and \texttt{MFPCA}, as documented by \cite{Happ_Greven_2018}.
\subsection{Data Generation}
To generate the functions $\{X^{(1)}_i, X^{(2)}_i, X^{(3)}_i\}$ for $i=1,\ldots,n$, we start by defining three mean functions and three covariance functions and six cross-covariance functions. The mean functions are defined as $\mu^{(1)}(t) = 5\sin(2\pi t)$, $\mu^{(2)}(t) = 5\cos(2\pi t)$, and $\mu^{(3)}(t) = 5(t - 1)^2$. Each function’s covariance structure is constructed using a set of orthonormal basis functions, we use $C_{kk}(s, t) = (\Phi^{(k)}(s))^\top \Lambda^{(k)} \Phi^{(k)}(t)$ for $k = 1, 2, 3$, where \(\Phi^{(k)}(t)\) contains three smooth basis functions
\[ \Phi^{(1)}(t) = \begin{bmatrix} \sqrt{2}\sin(2\pi t) \\ \sqrt{2}\cos(4\pi t) \\ \sqrt{2}\sin(4\pi t) \end{bmatrix}, \quad \Phi^{(2)}(t) = \begin{bmatrix} \sqrt{2}\cos(\pi t) \\ \sqrt{2}\cos(2\pi t) \\ \sqrt{2}\cos(3\pi t) \end{bmatrix}, \quad \Phi^{(3)}(t) = \begin{bmatrix} \sqrt{2}\sin(\pi t) \\ \sqrt{2}\sin(2\pi t) \\ \sqrt{2}\sin(3\pi t) \end{bmatrix},\]
and \(\Lambda^{(k)}\) is a diagonal matrix of eigenvalues
\[ \Lambda^{(1)} = \begin{pmatrix} 3 & 0 & 0 \\ 0 & 1.5 & 0 \\ 0 & 0 & 0.75 \end{pmatrix}, \quad \Lambda^{(2)} = \begin{pmatrix} 3.5 & 0 & 0 \\ 0 & 1.75 & 0 \\ 0 & 0 & 0.5 \end{pmatrix}, \quad \Lambda^{(3)} = \begin{pmatrix} 2.5 & 0 & 0 \\ 0 & 2 & 0 \\ 0 & 0 & 1 \end{pmatrix}. \]
Similarly, the cross-covariance functions between functions are defined as: 
\begin{equation*}
C_{kk'}(s, t) = \rho \Phi^{(k)}(s)^\top (\sqrt{\Lambda^{(k)}} \sqrt{\Lambda^{(k')}}) \Phi^{(k')}(t) \quad \text{for} \ k \ne k',
\end{equation*}
where \(\rho \in [0, 1]\) is a parameter that controls the overall level of correlation between the responses; if \(\rho = 0\), the responses are uncorrelated.

The eigendecomposition of the multivariate covariance function given by
$$
\left( \begin{array}{ccc}
C_{1,1} & C_{1,2} & C_{1,3} \\
C_{2,1} & C_{2,2} & C_{2,3}\\
C_{3,1} & C_{3,2} & C_{3,3} \end{array} \right)
$$
results in nine nonzero multivariate eigenvalues with associated multivariate eigenfunctions; hence, for $l = 1, \ldots, 9$, we simulate the scores $\eta_{il}$ from $N(0, d_l)$, where $d_l$ are the eigenvalues. The data is generated based on $Y_{i,j}^{(k)}=X_{i}^{(k)}(t^{(k)})+e_{ij}^{(k)},$ where $X_{i}^{(k)}(t^{(k)})$ is given in (\ref{U_Expan}) and
the measurement errors $e_{ij}^{(k)}$ were sampled independently from a normal distribution with mean zero and variance $\sigma_k^2$. The sampling observation points for each function are randomly selected from a uniform distribution within the unit interval. The number of observations for each subject, $m_{ik}$, is determined by a uniform discrete distribution over the set $\{3, 4, 5, 6, 7\}$. Consequently, the sampling points differ both between subjects and among functions within each subject.

We simulate datasets for six scenarios, each containing 100 replicates. The experimental design for the six scenarios is shown in Table \ref{Scenarios}. These simulated scenarios are also considered in the simulation study of \cite{Li_Xiao_Luo_2020}.
\begin{table}[htp!]
\centering
\begin{tabular}{lccccccc } 
 \hline
 Scenario &  1 &  2 &  3 &  4 &  5 &  6 \\
 \hline
 n & 25 & 100 & 500 & 25 & 100 & 500 \\ 
 $\sigma^2$ & 0.1 & 0.25 & 0.5 & 0.1 & 0.25 & 0.5 \\
 $\rho$ & 0.5 & 0.5 & 0.5 & 0.9 & 0.9 & 0.9 \\
 \hline
\end{tabular}
\caption{The parameters employed in the simulations to generate the six distinct scenarios.\label{Scenarios}}
\end{table}

To assess the performance of the methods, we measure the accuracy of the estimates by calculating the root mean squared error (RMSE) between the true values and estimates for the covariance function, eigenfunctions, and full curve estimation at 100 equally spaced points within the interval [0,1]. Additionally, we evaluate the relative squared error (RSE) between the true and estimated eigenvalues. Detailed descriptions of these performance metrics are provided in Appendix \ref{Sim}.

For the proposed method, we employ a B-Spline basis, selecting between 5 to 10 basis functions and possible ranks from 2 to 4 for each univariate FPC estimation. Optimal parameter values are determined using generalized cross-validation as described in Section \ref{FPCASp}. The number of multivariate principal components, denoted by \(M\), is determined using the elbow rule \citep{ahasverus}.

\subsection{The Covariance}

Table \ref{table:cov_est} provides the RMSE between the true and estimated covariance functions, evaluated at 100 equally spaced points within the interval [0,1] for each of the six simulated scenarios. The table compares performance across three methods: our proposed approach, \texttt{mFACEs}, and \texttt{MFPCA}. Our proposed method consistently achieves lower RMSEs than both \texttt{mFACEs} and \texttt{MFPCA} across all six simulated scenarios. 
\begin{table}[htp!]
\centering
\caption{Median and interquartile range (IQR) of the root mean squared error (RMSE) between the estimated covariance and the true covariance across the six simulated scenarios for the proposed method, \texttt{mFACEs}, and \texttt{MFPCA}. The smallest value for each column is highlighted in bold.} 
\label{table:cov_est}
 \begin{tabular}{lccc}

\hline

& \textbf{n = 25} & \textbf{n = 100} & \textbf{n = 500}  \\
\cline{2-4}
\textbf{Method}  &\multicolumn{3}{c}{$\mathbf{\rho = 0.5}$}\\
\hline
Proposed & \textbf{1.38 (0.26)} & \textbf{0.64 (0.13)} & \textbf{0.43} (0.07)\\
\texttt{mFACEs} & 1.75 (0.33) & 1.09 (0.18) & 0.59 (\textbf{0.06})\\
\texttt{MFPCA} & 1.69 (0.51) & 1.19 (1.73) & 0.87 (0.13)\\
\hline
&\multicolumn{3}{c}{$\mathbf{\rho = 0.9}$}\\

\hline
Proposed & \textbf{1.63 (0.44)} & \textbf{0.75} (0.21) & \textbf{0.59} (0.12)\\
\texttt{mFACEs} & 1.97 (\textbf{0.44}) & 1.17 (\textbf{0.20}) & 0.63 (\textbf{0.10})\\
\texttt{MFPCA} & 2.14 (0.51) & 1.62 (2.88) & 1.22 (0.17)\\
\hline
\end{tabular}
\end{table}

As expected, the accuracy for all methods improves as the sample size increases. Even under high correlation conditions ($\rho = 0.9$), our approach still maintains better performance, despite a slight decrease in accuracy. At a moderate correlation level \(\rho = 0.5\), we observe average percentage increases in RMSE of 45\% for \texttt{mFACEs} and 70\% for \texttt{MFPCA} compared to our method. Similarly, under stronger correlation \(\rho = 0.9\), the average percentage increase in RMSE relative to the proposed approach are 28\% and 85\% for \texttt{mFACEs} and \texttt{MFPCA}, respectively. Overall, the results highlight the advantage of our method in accurately recovering the covariance structure in sparse multivariate settings.

\subsection{The Eigenfunctions}

In Table \ref{table:rmse_psi}, we show the RMSE between the true and estimated top two eigenfunctions, evaluated at 100 equally spaced points within the interval [0,1] for each of the six simulated scenarios. The top two eigenvalues explain approximately 60\% of the total variation in the functional data when $\rho=0.5$ and 80\% when $\rho=0.9$.
\begin{table}[htp!]
\centering
\caption{Median and interquartile range (IQR) of the RMSEs (\(\times 100\)) for estimating the top two eigenfunctions, comparing our proposed method with \texttt{mFACEs} and \texttt{MFPCA} across six simulation scenarios. The smallest value for each column is highlighted in bold.} 
\label{table:rmse_psi}
 \resizebox{\textwidth}{!}{%
 \begin{tabular}{lcccccccc}
& & \multicolumn{3}{c}{\(\psi_1\)} && \multicolumn{3}{c}{\(\psi_2\)}\\
\cline{3-5}
\cline{7-9}
\textbf{Method} & & \textbf{n = 25} & \textbf{n = 100 } & \textbf{n = 500} & & \textbf{n = 25} & \textbf{n = 100 } & \textbf{n = 500 }\\

\cline{3-9}
&&\multicolumn{7}{c}{$\mathbf{\rho = 0.5}$}\\
\hline
Proposed &&\textbf{ 27.20} (21.49) & \textbf{12.55} (6.92) & \textbf{5.89} (3.10) &&\textbf{ 41.61 (23.64)} & \textbf{16.90 (8.11)} & \textbf{9.25 (3.32)}\\
\texttt{mFACEs} && 27.47 (\textbf{12.58}) & 15.71 (\textbf{6.32}) & 7.94 (3.06) && 58.04 (23.67) & 30.94 (16.55) & 14.24 (3.67)\\
\texttt{MFPCA} && 32.62 (28.18) & 16.68 (23.90) & 7.30 (\textbf{3.05}) && 58.25 (20.65) & 32.29 (24.55) & 20.62 (3.75)\\
\hline
&&\multicolumn{7}{c}{$\mathbf{\rho = 0.9}$}\\

\hline
Proposed && \textbf{20.06} (22.64) & \textbf{9.87 (7.61)} & \textbf{4.80} (3.96) && \textbf{32.35} (21.80) & \textbf{12.60} (6.80) & \textbf{6.51} (3.32)\\
\texttt{mFACEs} && 20.85 (\textbf{15.26}) & 12.62 (8.00) & 7.02 (\textbf{3.53}) && 38.52 (\textbf{17.34}) & 18.65 (\textbf{5.54}) & 10.68 (3.21)\\
\texttt{MFPCA} && 24.38 (30.00) & 13.99 (30.07) & 6.27 (2.50) && 42.90 (19.06) & 22.64 (34.46) & 15.48 (\textbf{1.97})\\
\hline
\end{tabular}}   
\end{table}
 
The proposed approach consistently outperforms the existing methods in terms of accuracy across both eigenfunctions and all six scenarios.
Compared to the proposed approach, the average percentage increase in RMSE for \(n=25\) is 15\% with \texttt{mFACEs} and 28\% with \texttt{MFPCA}. For \(n=100\), the increase is 46\% with \texttt{mFACEs} and 61\% with \texttt{MFPCA}. At \(n=500\), \texttt{mFACEs} shows a 49\% increase in RMSE, while \texttt{MFPCA} shows a 78\% increase. Across all methods, the accuracy increases with the sample size.

\subsection{The Eigenvalues} 
Table \ref{table:rse_eta} summarizes the performance of the proposed method, \texttt{mFACEs}, and \texttt{MFPCA} in estimating the top two eigenvalues (\(\eta_2\) and \(\eta_2\)) across the six simulated scenarios. The table presents the median and IQR of the relative squared error.

\begin{table}[htp!]
\centering
\caption{100 \(\times\) Median and interquartile range (IQR) of the relative squared error for the top two eigenvalues, estimated using our proposed method, \texttt{mFACEs}, and \texttt{MFPCA} across six different simulated scenarios. The smallest value for each column is highlighted in bold.} 
\label{table:rse_eta}
 \resizebox{\textwidth}{!}{%
 \begin{tabular}{lcccccccc}
& & \multicolumn{3}{c}{\(\eta_1\)} && \multicolumn{3}{c}{\(\eta_2\)}\\
\cline{3-5}
\cline{7-9}
\textbf{Method} & & \textbf{n = 25} & \textbf{n = 100 } & \textbf{n = 500} & & \textbf{n = 25} & \textbf{n = 100 } & \textbf{n = 500 }\\

\cline{3-9}&&\multicolumn{7}{c}{$\mathbf{\rho = 0.5}$}\\
\hline
Proposed && \textbf{2.41 (7.49)} &\textbf{ 1.01 (2.73)} & 1.05 (1.37) && \textbf{3.55 (6.10)} & \textbf{1.07 (2.20)} & 2.33 (2.76)\\
\texttt{mFACEs} && 3.58 (10.91) & 1.55 (3.82) & \textbf{0.32 (0.73)} && 4.53 (9.75) & 3.66 (7.84) & \textbf{0.80 (1.30)}\\
\texttt{MFPCA} && 11.86 (20.28) & 7.64 (14.54) & 4.00 (3.41) && 19.49 (26.16) & 18.50 (17.2) & 13.95 (6.40)\\
\hline
&&\multicolumn{7}{c}{$\mathbf{\rho = 0.9}$}\\

\hline
Proposed &&\textbf{ 2.61 (7.37)} & \textbf{1.32 (2.26)} & 1.22 (1.70) &&\textbf{6.62 (12.60)} & \textbf{1.80 (3.04)} & 3.07 (2.95)\\
\texttt{mFACEs} && 2.94 (8.48) & 1.43 (3.93) & \textbf{0.37 (0.76)} &&7.56 (16.18) & 2.97 (5.74) & \textbf{0.66 (1.14)}\\
\texttt{MFPCA} && 9.43 (21.60) & 7.16 (19.26) & 5.14 (4.17) &&30.48 (39.19) & 23.06 (20.92) & 17.93 (8.04)\\
\hline
\end{tabular}}   
\end{table}

The proposed method consistently performs better for small and moderate sample sizes ($n = 25, \ 100$), consistently showing the lowest median error and IQR across all scenarios.  In contrast, \texttt{mFACEs} generally performs better at large sample sizes ($n = 500$), yielding the smallest errors in most cases for both \(\eta_2\) and \(\eta_2\). The results also reveal that higher correlation (\(\rho = 0.9\)) generally increases estimation difficulty across all methods, as evidenced by larger median errors and interquartile ranges compared to the moderate correlation scenario \(\rho = 0.5\). 

\subsection{The Curve Reconstruction} 
Table \ref{table:rse_xhat} presents the median and interquartile range (IQR) of the Root Mean Squared Error (RMSE) between the true underlying curves and their estimates for three methods across 100 equally spaced evaluation points. 
\begin{table}[h!]
\centering
\caption{Median and interquartile range (IQR) of the Root Mean Squared Error between the true curves and their estimates evaluated at 100 equally spaced points within the interval $[0,1]$ for all three methods across each of the six simulated scenarios. The smallest value for each column is highlighted in bold.} 
\label{table:rse_xhat}
 \begin{tabular}{lcccc}
\hline
\textbf{Method} && \textbf{n = 25} & \textbf{n = 100} & \textbf{n = 500} \\
\hline
&&\multicolumn{3}{c}{$\mathbf{\rho = 0.5}$}\\
\hline
Proposed && \textbf{0.99 (0.25)} & \textbf{0.67 (0.06)} & \textbf{0.75 (0.03)}\\
\texttt{mFACEs} && 1.44 (0.79) & 1.10 (0.14) & 0.89 (0.06)\\
\texttt{MFPCA} && 1.40 (0.51) & 1.09 (0.88) & 0.97 (0.04)\\
\hline
&&\multicolumn{3}{c}{$\mathbf{\rho = 0.9}$}\\
\hline
Proposed && \textbf{1.00 (0.23)} & \textbf{0.66 (0.07)} & 0.75 (\textbf{0.03})\\
\texttt{mFACEs} && 1.17 (0.24) & 0.85 (0.08) & \textbf{0.70 (0.03)}\\
\texttt{MFPCA} && 1.34 (0.37) & 1.10 (1.15) & 0.97 (0.05)\\
\hline
\end{tabular}
\end{table}

The proposed method demonstrates better or similar performance across all scenarios, showing the lowest median RMSE values for small to moderate sample sizes ($n = 25, 100$). Its estimation quality remains stable showing minimal variability (smallest Interquartile Ranges) regardless of correlation levels ($\rho = 0.5$ vs $0.9$). When the sample size is large ($n=500$) and the correlation is high ($\rho=0.9$), \texttt{mFACEs} exhibits marginally superior performance relative to the proposed method. This difference is consistent with enhanced accuracy in eigenvalue estimation by \texttt{mFACEs} under these conditions, indicating that its estimates benefit from the increased sample size.

\section{Alzheimer’s disease}\label{app:Alzheimer}

The proposed method was applied to data from the Alzheimer's Disease Neuroimaging Initiative (ADNI), a longitudinal multicenter study aimed at early detection of Alzheimer's disease through biomarkers, imaging, clinical, and neuropsychological assessments \citep{Weiner_2017}. The initial phase, ADNI-1, began in 2004 and involved 402 patients with mild cognitive impairment (MCI), considered a risk state for Alzheimer's disease. Subsequent phases, ADNIGO and ADNI-2, added 582 patients with MCI and significant memory concerns. ADNI-3 further included 485 MCI patients, comprising both early and late stages of MCI. These figures are sourced from the merged dataset (ADNIMERGE) in the publicly available ADNI R package, accessible at \url{https://ida.loni.usc.edu}.

The objectives of our analysis are to estimate the overall trend in cognitive decline over time, extract the dominant modes of variation, and determine how individual subjects deviate from the overall trend. For our analysis, we selected patients who had at least one follow-up visit, resulting in a sample of 929 individuals. We included assessments from baseline (month 0) up to 120 months, covering data from ADNI-1 to ADNI-3. In line with findings from \cite{Li_Chan_Doody_Quinn_Luo_2017}, which showed that cognitive measures are more predictive of Alzheimer's disease progression than imaging, we focused on the top three cognitive metrics: the Alzheimer's Disease Assessment Scale-Cognitive 13 items (ADAS-Cog 13), Rey Auditory Verbal Learning Test immediate recall (RAVLT.imme), and Rey Auditory Verbal Learning Test learning curve (RAVLT.learn).  The ADAS-Cog 13 consists of 13 tasks that evaluate memory, language skills, praxis, attention, and other cognitive abilities. A high score on the ADAS-Cog 13 indicates greater cognitive impairment. For RAVLT.imme a high score generally indicates better short-term and working memory abilities. RAVLT.learn measures an individual's ability to learn and retain new verbal information over multiple trials. Lower RAVLT.learn scores typically indicate greater impairment in the ability to learn and remember new information. We excluded the Alzheimer's Disease Assessment Scale-Cognitive 11 items (ADAS-Cog 11) due to its high correlation (correlation coefficient of .98) with ADAS-Cog 13. Due to skewness in the ADAS-Cog 13 data, we applied a square root transformation to these observations.

\begin{figure}[htp!]
    \centering
    \includegraphics[width=\textwidth]{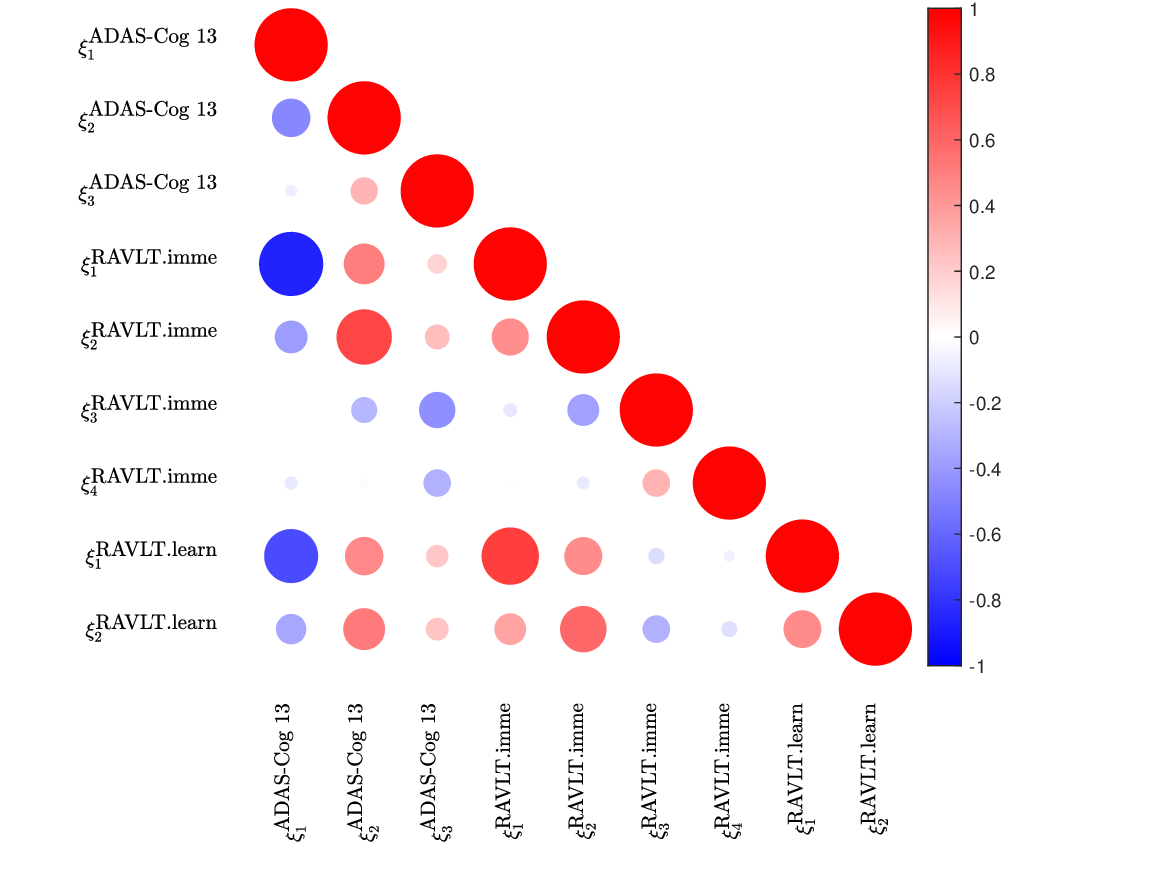}
    \caption{The empirical correlation between the univariate functional principal component scores.}
    \label{fig:app_scores_Corr}
\end{figure}

Figure \ref{fig:app_scores_Corr} illustrates the empirical correlation between the univariate functional principal component scores.
As illustrated in Figure \ref{fig:app_scores_Corr}, there are moderate to high correlations between the principal component scores. The second principal component score for RAVLT.imme correlates at 0.72 with the second principal component score for ADAS-Cog 13. The first principal component score for RAVLT.imme correlates at 0.75 with the first principal component score for RAVLT.learn. Furthermore, the first principal component score for RAVLT.imme exhibits a negative correlation of -0.86 with the first principal component score for ADAS-Cog 13. 

The first two principal components account for 98\% of the total variation in the multivariate functional data: the first component explains 92\%, and the second 6\%. 

\begin{figure}[htp!]
    \centering
    \includegraphics[width=\textwidth]{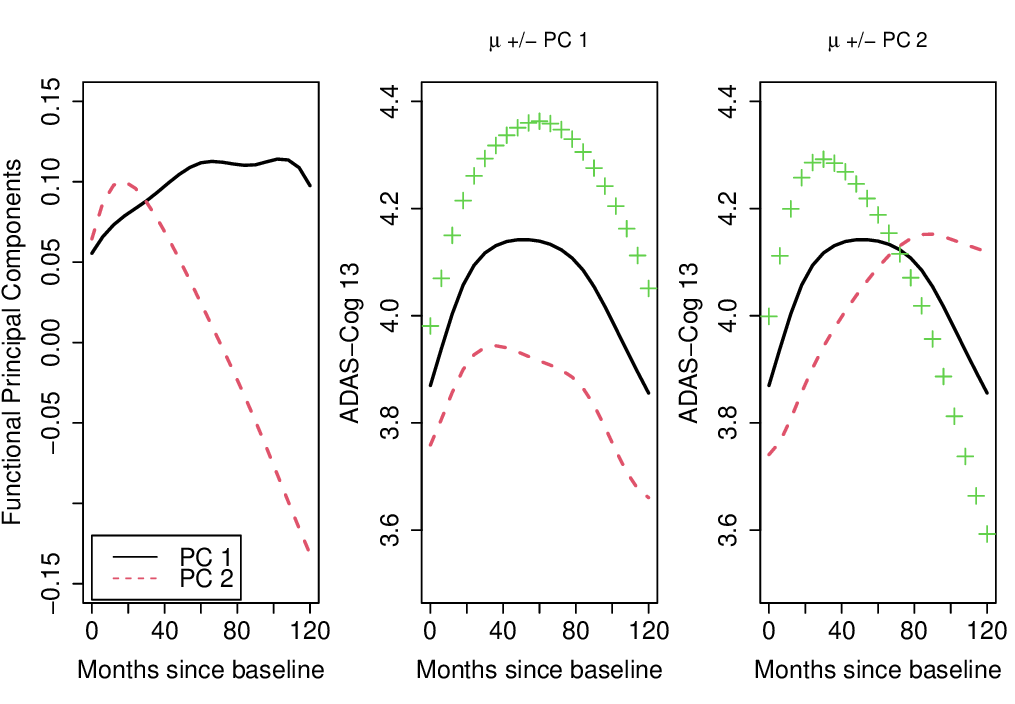}
    \caption{Estimated top two eigenfunctions for the longitudinal marker ADAS-Cog 13 (Disease Assessment Scale-Cognitive 13 items). In the two right-hand panels, the black solid curve represents the overall smoothed mean, which is identical across all cases. The remaining curves show the effect of adding (green, “+”) or subtracting (red, “– –”) an appropriately scaled multiple of the corresponding principal component.}
    \label{fig:app_eigFunc_est_1}
\end{figure}

Figure \ref{fig:app_eigFunc_est_1} illustrates the first two functional principal components for ADAS-Cog 13 and their effects when added to and subtracted from the mean ADAS-Cog 13 values. The black curve represents the overall smoothed mean, which remains the same in all cases. The other two curves depict the effect of adding (green, “+”) and subtracting (red, “– –”) an appropriate multiple of the respective principal component.
The black curve in Figure \ref{fig:app_eigFunc_est_1} illustrates the overall smoothed mean of ADAS-Cog 13 scores, displaying a concave shape with a peak at 50 months after the initial assessment. The first eigenfunction primarily captures the overall level of the ADAS-Cog 13 scores relative to the data's mean trajectory. Individuals with high scores on the first principal component have above-average ADAS-Cog 13 scores, which indicates more severe cognitive decline than the average observed in the dataset. Conversely, subjects with low scores on this component typically show below-average ADAS-Cog 13 scores, suggesting a less severe cognitive impairment over time. The second eigenfunction is focused on the linear trend in the ADAS-Cog 13 scores over time. Subjects with high scores on this second principal component exhibit a decline in their ADAS-Cog 13 scores, indicating they are improving their cognitive function. In contrast, subjects with low (or negative) scores on this component have increasing ADAS-Cog 13 scores over time, which points to a worsening cognitive state. The second eigenfunction, therefore, effectively captures the direction of ADAS-Cog 13 scores over time, indicating whether a subject's cognitive function is declining or improving.

\begin{figure}[htp!]
    \centering
    \includegraphics[width=\textwidth]{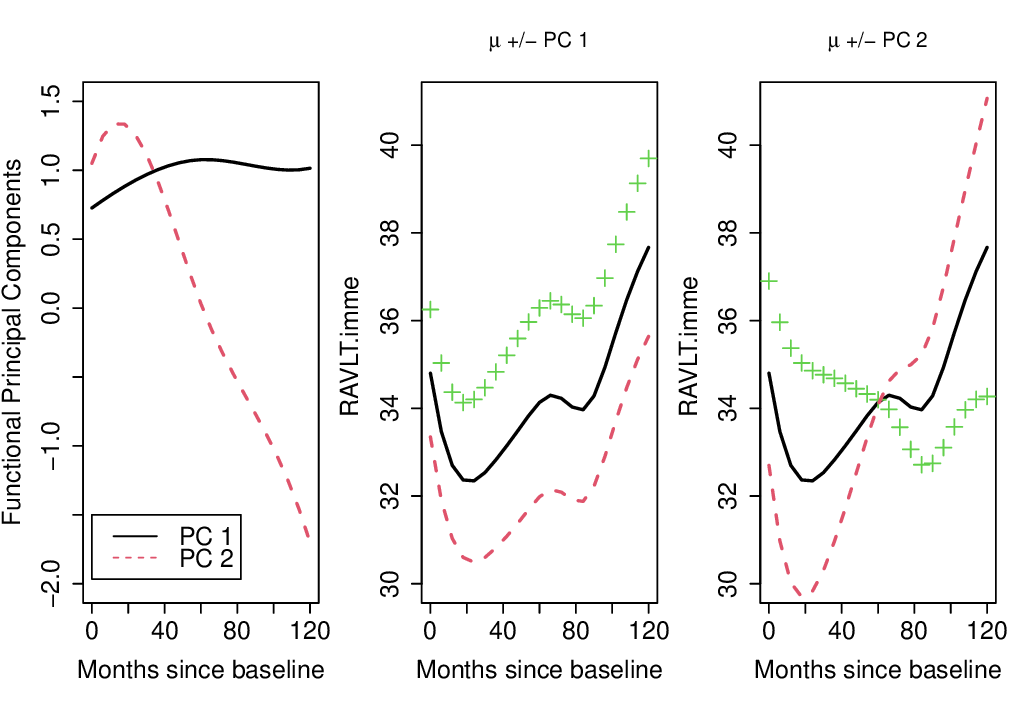}
    \caption{Estimated top two eigenfunctions for the longitudinal marker Rey Auditory Verbal Learning Test immediate recall (RAVLT.imme). In the two right-hand panels, the black solid curve represents the overall smoothed mean, which is identical across all cases. The remaining curves show the effect of adding (green, “+”) or subtracting (red, “– –”) an appropriately scaled multiple of the corresponding principal component.}
    \label{fig:app_eigFunc_est_2}
\end{figure}

 Figure \ref{fig:app_eigFunc_est_2} illustrates the first two functional principal components for RAVLT.imme and their effects when added to and subtracted from the mean RAVLT.imme values. In the two right panels, the black solid curve denotes the overall smoothed mean, which is identical across all cases. The other two curves illustrate the effect of adding (red, “+”) and subtracting (yellow, “– –”) a suitably scaled multiple of the corresponding principal component. The first principal component primarily captures vertical shifts from the mean RAVLT.imme trajectory. Individuals with a positive score on this component consistently have above-average RAVLT.imme scores over time, which suggests better-than-average short-term and working memory abilities. Conversely, those with negative scores on this component show below-average RAVLT.imme scores, indicating poorer memory performance. The second principal component appears to capture the change in RAVLT.imme scores over time. Subjects with a positive score on this component show a decrease in their RAVLT.imme scores, indicating a worsening of their short-term and working memory. In contrast, those with a negative score show an increase in their RAVLT.imme scores, implying an improvement in their short-term and working memory.

\begin{figure}[htp!]
    \centering
    \includegraphics[width=\textwidth]{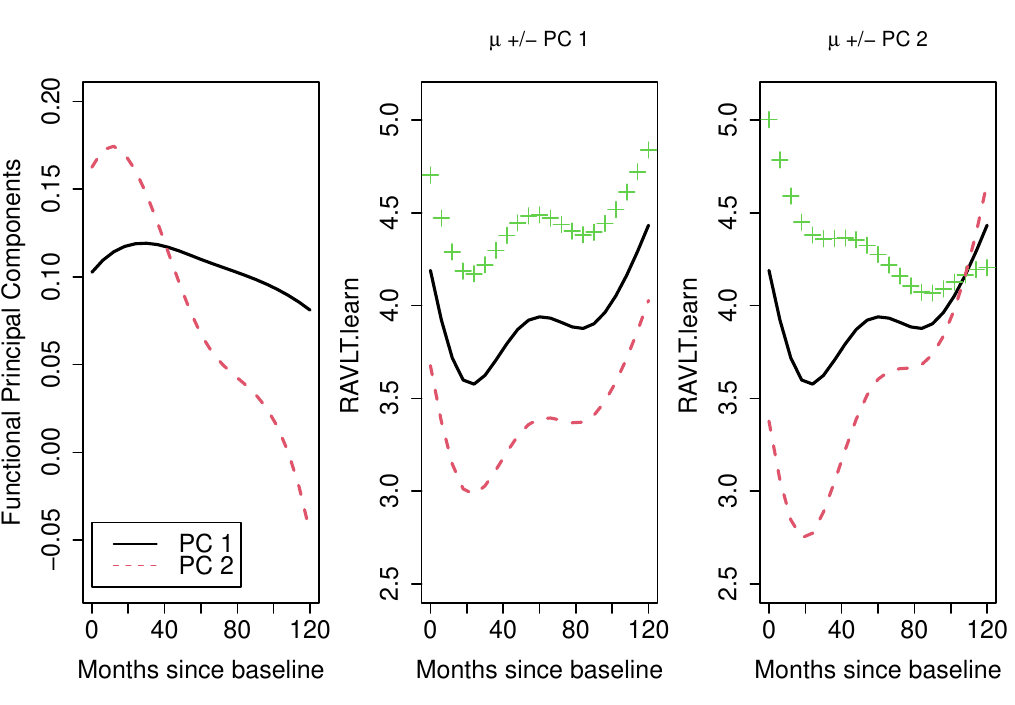}
    \caption{Estimated top two eigenfunctions for the longitudinal marker Rey Auditory Verbal Learning Test immediate recall (RAVLT.learn). In the two right-hand panels, the black solid curve represents the overall smoothed mean, which is identical across all cases. The remaining curves show the effect of adding (green, “+”) or subtracting (red, “– –”) an appropriately scaled multiple of the corresponding principal component.}
    \label{fig:app_eigFunc_est_3}
\end{figure}

Figure \ref{fig:app_eigFunc_est_3} illustrates the first two functional principal components for RAVLT.learn and their effects when added to and subtracted from the mean RAVLT.learn values. The black curve represents the overall smoothed mean, which remains the same in all cases. The other two curves show the effect of adding (green, “+”) and subtracting (red, “– –”) an appropriate multiple of the respective principal component. The first principal component captures vertical deviations of RAVLT.learning scores from the data average. Individuals with higher scores on this component have above-average RAVLT.learning scores, which indicates a better ability to learn and retain new information. The second principal component appears to capture linear trend in the RAVLT.learning scores over time among subjects. Individuals with positive scores on this component exhibit a declining trend in RAVLT learning scores, indicating diminished ability to learn and retain new information, whereas negative scores reflect the opposite pattern.

\begin{figure}[htp!]
    \centering
    \includegraphics[width=\textwidth]{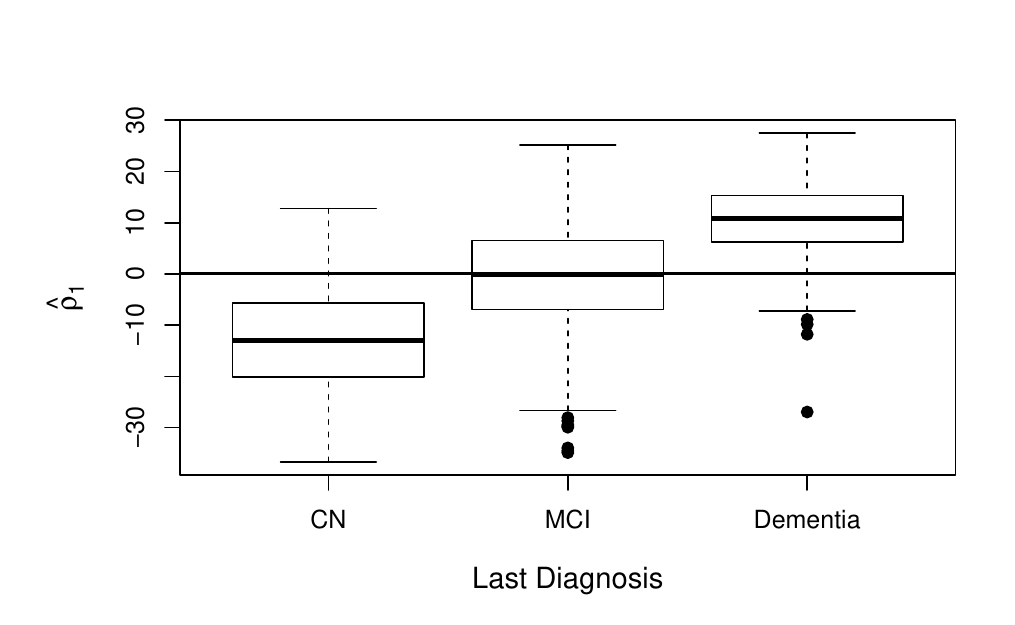}
    \caption{Estimated scores vs the last diagnosis of patients.}
    \label{fig:app_scores_diagnosis}
\end{figure}

Figure \ref{fig:app_scores_diagnosis} shows the relationship between the first multivariate principal scores and the last diagnosis for the subjects. As expected, subjects with mainly positive scores are those diagnosed with dementia by their last visit, while scores of subjects who remained cognitively normal (CN) during their last visit are nearly all negative. Persons with mild cognitive impairment (MCI) have intermediate score values. 

\section{Milk Quality in Dairy Cattle}\label{app:Milk}

Dairy cows experience a lactation cycle starting after calving, during which they produce milk. Somatic cells, which are white blood cells present in milk, are essential for combating infection and repairing tissue. An increase in these cells occurs when a cow's udder is infected, migrating into the milk to fight bacteria. Consequently, the somatic cell count (SCC) in milk serves as a crucial indicator of udder health and milk quality. In addition, poor udder health status can have a detrimental effect on milk yield, leading to reductions in the dairy farm's profit. Our study utilizes a dataset from Irish research dairy farms, provided by the VistaMilk SFI Research Centre, which records SCC and milk yields over a 44-week milking period beginning five days post-calving. The objectives of our analysis are to estimate the overall trend in SCC and milk yield in cows over time, extract dominant modes of variation, and determine how individual cows' levels deviate from the overall trend. 

\begin{figure}[htp!]
    \centering
    \includegraphics[width=\textwidth]{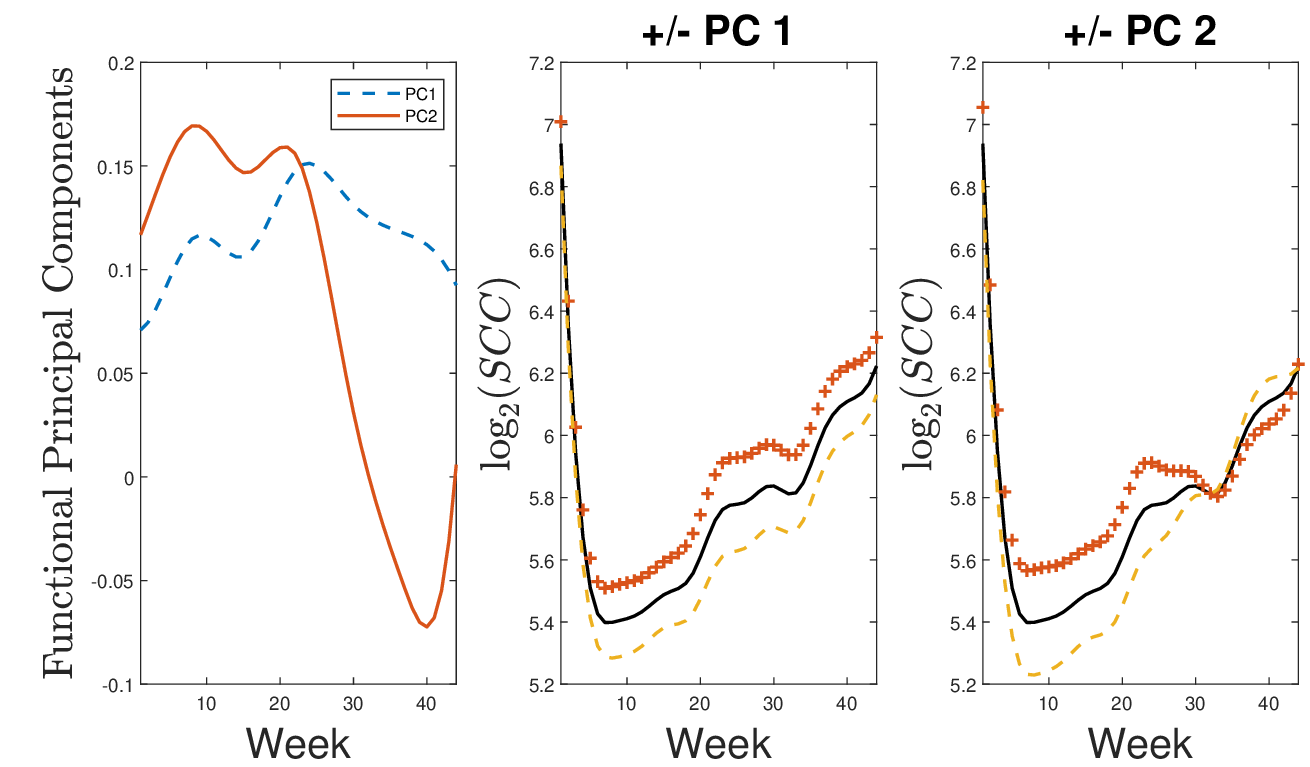}
    \caption{Estimated top two eigenfunctions for $\log_{2}SCC$. In the two right-hand panels, the black solid curve represents the overall smoothed mean, which is identical across all cases. The remaining curves show the effect of adding (red, “+”) or subtracting (yellow, “– –”) an appropriately scaled multiple of the corresponding principal component.}
    \label{fig:app_eigFunc_est_1_VM}
\end{figure}

The first two principal components account for 93\% of the total variation in the multivariate functional data: the first component explains 78\% and the second 15\%. Figure \ref{fig:app_eigFunc_est_1_VM} illustrates the first two multivariate functional principal components and their effects when added to and subtracted from the mean $\log_{2}SCC$. The black curve represents the overall smoothed mean, which remains the same in all cases. The other two curves show the effect of adding (red curve with crosses) and subtracting (yellow dashed curve) an appropriate multiple of the respective principal component.
The mean curve shows an initial decline from week 1 to week 9, a gradual rise from weeks 9 to 19, a sharp increase from weeks 19 to 24, stabilization from weeks 24 to 33, and a subsequent rise thereafter. The first eigenfunction primarily reflects the level in SCC scores among cows. Cows with low scores on the first principal component typically exhibit below-average SCC scores, while those with high scores display above-average values. The second eigenfunction appears to distinguish between cows exhibiting a rapid decrease in SCC scores followed by a rapid increase (yellow dashed curve), and those exhibiting a gradual decrease followed by a gradual increase (red curve with crosses).

\begin{figure}[htp!]
    \centering
    \includegraphics[width=\textwidth]{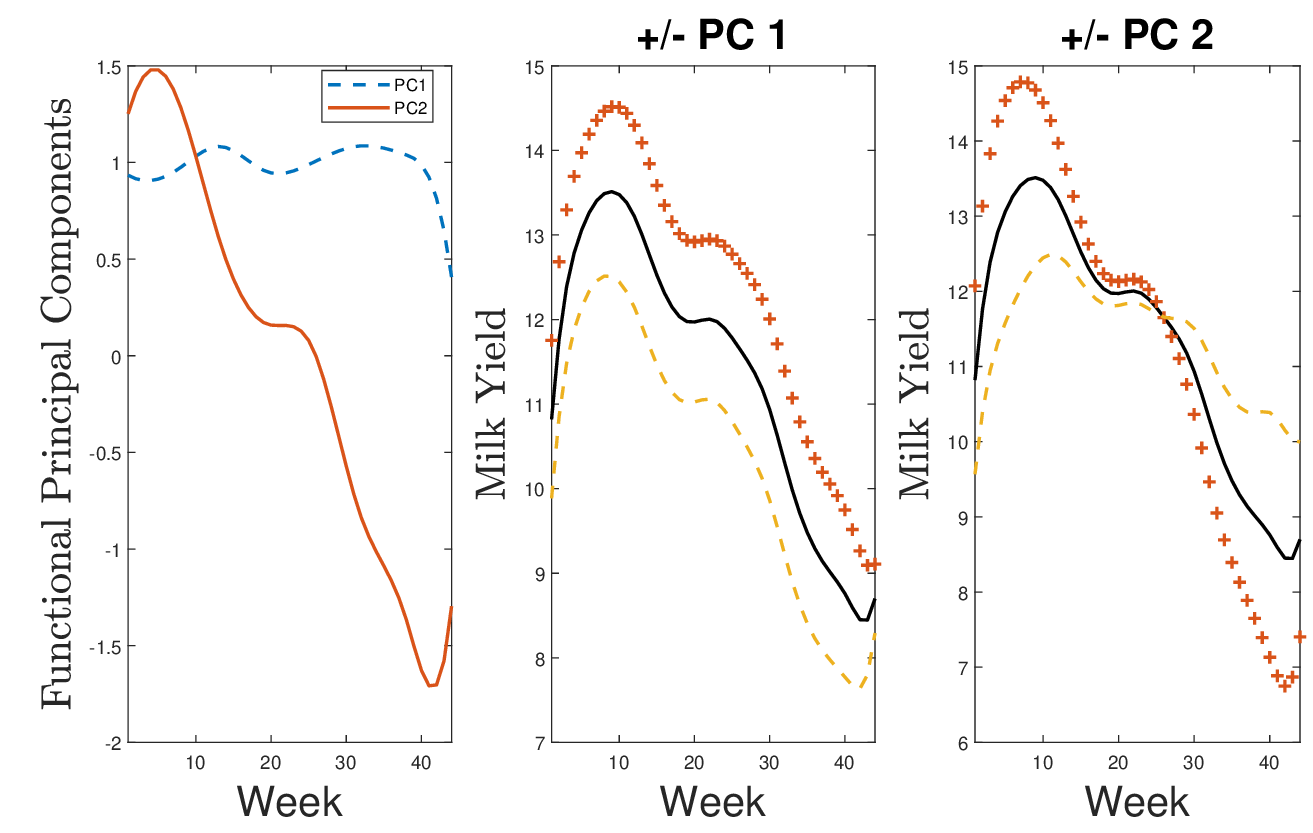}
    \caption{Estimated top two eigenfunctions for Milk Yield. In the two right-hand panels, the black solid curve represents the overall smoothed mean, which is identical across all cases. The remaining curves show the effect of adding (red, “+”) or subtracting (yellow, “– –”) an appropriately scaled multiple of the corresponding principal component.}
    \label{fig:app_eigFunc_est_2_VM}
\end{figure}

Figure \ref{fig:app_eigFunc_est_2_VM} illustrates the first two multivariate functional principal components and their effects when added to and subtracted from the mean milk yield. The black curve represents the overall smoothed mean, which remains the same in all cases. The other two curves show the effect of adding (red curve with crosses) and subtracting (yellow dashed curve) an appropriate multiple of the respective principal component.
The smoothed mean trajectory clearly illustrates the typical lactation curve seen in dairy production. Average milk yield begins at approximately 11 liters around week 5, steadily increasing to a peak at week 9, then gradually declining thereafter. This mirrors the classic lactation pattern: a rapid rise in milk production post-calving, reaching a peak, followed by a sustained decline over the remainder of the lactation period.

The first eigenfunction predominantly reflects the overall level of milk yield among cows. Cows with low scores on this component generally show below-average milk production, while those with high scores exhibit above-average yields. The second eigenfunction, however, appears to capture differences in the shape of the lactation curve. It distinguishes between cows with a sharp decline after their peak period (represented by the red curve with crosses) and those showing a slow decline in production after their peak period (shown by the yellow dashed curve).

\section{Conclusion}

We propose a novel method for Functional Principal Component Analysis designed specifically for multivariate sparse functional data. This method extends the univariate FPCA framework for sparse data, as detailed in \cite{Mbaka_Cao_Carey_2024}, to accommodate multivariate observations. It ensures orthogonality in the basis function expansion of the eigenfunctions and guarantees positive eigenvalues, resulting in a low-rank, positive semi-definite covariance function. Simulation studies highlight the substantial advancements of our method. Across all six scenarios, covariance estimation accuracy improved by an average of 36\% compared to \texttt{mFACEs} and 77\% compared to \texttt{MFPCA}. Similarly, the estimation accuracy of full trajectories had an average improvement of 28\% over \texttt{mFACEs} and 44\% over \texttt{MFPCA}. These figures represent the average percentage decrease in RMSE of our proposed method compared to the competing methods, underscoring our method's consistent performance gains.
Our approach has been applied effectively to the Alzheimer's Disease Neuroimaging Initiative (ADNI), a longitudinal multicenter study focused on the early detection of Alzheimer's disease. The patterns observed in the multivariate functional principal components align with a similar analysis performed on this dataset as shown in \cite{Li_Xiao_Luo_2020}. In addition we analyze the main modes of variations in a dataset containing somatic cell scores and milk yield in dairy cattle.

Further research could explore incorporating covariate effects into the analysis, as examined in \cite{chiou2003functional}, \cite{cardot2007conditional}, and \cite{jiang2009covariate}. Our method can be adapted to scenarios where eigenvalues are modeled as simple parametric functions of a covariate. In this framework, the eigenfunctions can remain independent or become dependent on the covariate by incorporating these covariates into the basis function expansion of \(u_{q}^{(k)}\).

FPCA is extensively utilized to identify major sources of variation in samples of curves. These sources are captured by the functional principal components which are typically non-zero across the entire domain and can be complex to interpret. Recent research by \cite{nie2020sparse} has focused on estimating sparse functional principal components that are non-zero only in specific subregions, aiding in both the identification of major variation sources and the localization of these variations within specific subregions. Our method could potentially be extended to estimate these sparse functional principal components.

\section{Acknowledgement}
 We are grateful to the Associate Editor and the two referees for their insightful and helpful comments, which have greatly improved the quality of this work. We also acknowledge the financial support of Research Ireland (SFI) and the Department of Agriculture, Food and Marine on behalf of the Government of Ireland under Grant Number [16/RC/3835] - VistaMilk and Research Ireland (SFI).

\section{Funding details} This work was supported by Research Ireland (SFI) and the Department of Agriculture, Food and Marine on behalf of the Government of Ireland under Grant Number [16/RC/3835] - VistaMilk and Research Ireland (SFI).

\section{Disclosure statement} The authors report there are no competing interests to declare.

\section{Data availability statement} 
Reproducible examples and the R code for this methodology, are available at \\ \url{https://github.com/uchembaka/mGSFPCA}.
\newpage

\section{Appendix}

\subsection{The weighted version}\label{Weighted}

When the functions exhibit significant differences in domains, ranges, or amounts of variation, weights are necessary to derive multivariate functional principal components that offer meaningful interpretations \citep{Chiou_Yang_Chen_2014}. For such cases, the weighted inner product is defined as \(\langle \mathbf{f}, \mathbf{g} \rangle_{\mathcal{H}}^{w} = \sum_{k=1}^p w_k \langle f^{(k)}, g^{(k)} \rangle_{\mathcal{H}}\), where \(w_k\) is a positive weight associated with the \(k\)th function. 

The corresponding weighted covariance operator $\Gamma_{w} f$ is given by its elements $(\Gamma_{w} f)^{(k')}$ with $f \in \mathcal{H}$ and $(\Gamma_{w} f)^{(k')}(t^{(k')}) = \left<C_{.k'}(.,t^{(k')}),\mathbf{f}\right>_{\mathcal{H}}^{w},$ for $t^{(k')} \in \mathcal{T}^{(k')}.$

\subsection{Simulations}\label{Sim}

\begin{itemize}
    \item \textbf{Root Mean Squared Error (RMSE) of the Covariance Estimate:}
    \[
    \text{RMSE}_{C} = \sqrt{\frac{1}{|\mathcal{G}|^2} \sum_{h \in \mathcal{G}} \sum_{h' \in \mathcal{G}} (\hat{C}(h,h') - C(h,h'))^2}
    \]

    \item \textbf{Root Mean Squared Error (RMSE) of the Eigenfunction and Score Estimate:}
    \[
    \text{RMSE}_{\psi_l} = min\left[\sqrt{\frac{1}{|\mathcal{G}|} \sum_{h \in \mathcal{G}} (\hat{\psi}_l(h) - \psi_l(h))^2}, \sqrt{\frac{1}{|\mathcal{G}|} \sum_{h \in \mathcal{G}} (\hat{\psi}_l(h) + \psi_l(h))^2}\right] \]
    \item \textbf{Relative Squared Error (RSE) of the Eigenvalues:}
    \[
    RSE_{\eta_l} = \frac{(\hat{\eta}_l - \eta_l)^2}{\eta_l^2};
    \]
    \item \textbf{Root Mean Squared Error (RMSE) of Curve Reconstruction:}
    \[
    \text{RMSE}_{\hat{X}^c} = \sqrt{\frac{1}{n|\mathcal{G}|} \sum^n_{i=1} \sum_{h \in \mathcal{G}} (\hat{X}^c_i(h) - X^c_i(h))^2},
    \]
    where is the centered function $\hat{X}^c = \sum^M_{l=1} \hat{\rho}_l \hat{\psi}_l(\mathbf{t})$.
\end{itemize}

\newpage

\bibliographystyle{apalike}
\bibliography{ref.bib}

@article{mercer1909xvi,
  title={Xvi. functions of positive and negative type, and their connection the theory of integral equations},
  author={Mercer, James},
  journal={Philosophical transactions of the royal society of London. Series A, containing papers of a mathematical or physical character},
  volume={209},
  number={441-458},
  pages={415--446},
  year={1909},
  publisher={The Royal Society London}
}

@article{loeve1945fonctions,
  title={Sur les fonctions al{\'e}atoires stationnaires du second ordre},
  author={Lo{\`e}ve, Michel},
  journal={Revue Scientifique},
  volume={83},
  pages={297--303},
  year={1945}
}

@article{karhunen1947under,
  title={Under lineare methoden in der wahr scheinlichkeitsrechnung},
  author={Karhunen, Kari},
  journal={Annales Academiae Scientiarun Fennicae Series A1: Mathematia Physica},
  volume={47},
  year={1947}
}

@article{reed1980methods,
  title={Methods of Modern Mathematical Physics I: Functional Analysis, Acad},
  author={Reed, M and Simon, B},
  journal={Press, San Diego},
  year={1980}
}

@book{Ramsay_Silverman_2005,
  address={New York},
  edition={2nd ed},
  series={Springer series in statistics},
  title={Functional Data Analysis},
  ISBN={9780387400808},
  callNumber={QA278 .R36 2005},
  publisher={Springer},
  author={Ramsay, J. O. and Silverman, B. W.},
  year={2005},
  collection={Springer series in statistics}
}

@book{jiang2009covariate,
  title={Covariate adjusted functional principal component analysis},
  author={Jiang, Ci-Ren},
  year={2009},
  publisher={University of California, Davis}
}

@article{saporta1981methodes,
  title={M{\'e}thodes exploratoires d'analyse de donn{\'e}es temporelles},
  author={Saporta, Gilbert},
  journal={Cahiers du Bureau universitaire de recherche op{\'e}rationnelle S{\'e}rie Recherche},
  volume={37},
  pages={7--194},
  year={1981}
}

@article{Dauxois_Pousse_Romain_1982,
  title={Asymptotic theory for the principal component analysis of a vector random function: Some applications to statistical inference},
  volume={12},
  rights={https://www.elsevier.com/tdm/userlicense/1.0/},
  ISSN={0047259X},
  url={https://linkinghub.elsevier.com/retrieve/pii/0047259X82900884},
  DOI={10.1016/0047-259X(82)90088-4},
  number={1},
  journal={Journal of Multivariate Analysis},
  author={Dauxois, J. and Pousse, A. and Romain, Y.},
  year={1982},
  month=mar,
  pages={136–154},
  language={en}
}

@article{Ramsay_Dalzell_1991,
  title={Some Tools for Functional Data Analysis},
  volume={53},
  ISSN={00359246},
  url={https://onlinelibrary.wiley.com/doi/10.1111/j.2517-6161.1991.tb01844.x},
  DOI={10.1111/j.2517-6161.1991.tb01844.x},
  number={3},
  journal={Journal of the Royal Statistical Society: Series B (Methodological)},
  author={Ramsay, J. O. and Dalzell, C. J.},
  year={1991},
  month={Jul},
  pages={539–561},
  language={en}
}

@article{Silverman_1996,
    title={Smoothed functional principal components analysis by choice of norm},
    volume={24},
    ISSN={0090-5364},
    url={https://projecteuclid.org/journals/annals-of-statistics/volume-24/issue-1/Smoothed-functional-principal-components-analysis-by-choice-of-norm/10.1214/aos/1033066196.full},
    DOI={10.1214/aos/1033066196},
    number={1},
    journal={The Annals of Statistics},
    author={Silverman, Bernard W.},
    year={1996},
    month=feb
}

@article{Hall_Müller_Wang_2006,
  title={Properties of principal component methods for functional and longitudinal data analysis},
  volume={34},
  ISSN={0090-5364},
  url={https://projecteuclid.org/journals/annals-of-statistics/volume-34/issue-3/Properties-of-principal-component-methods-for-functional-and-longitudinal-data/10.1214/009053606000000272.full},
  DOI={10.1214/009053606000000272},
  number={3},
  journal={The Annals of Statistics},
  author={Hall, Peter and Müller, Hans-Georg and Wang, Jane-Ling},
  year={2006},
  month={Jun}
}

@article{Wong_Zhang_2019,
    title={Nonparametric operator-regularized covariance function estimation for functional data},
    volume={131},
    ISSN={01679473},
    url={https://linkinghub.elsevier.com/retrieve/pii/S0167947318301221},
    DOI={10.1016/j.csda.2018.05.013},
    journal={Computational Statistics \& Data Analysis},
    author={Wong, Raymond K.W. and Zhang, Xiaoke},
    year={2019},
    month=mar,
    pages={131–144},
    language={en}
}

@article{Shi_Weiss_Taylor_1996,
    title={An Analysis of Paediatric CD4 Counts for Acquired Immune Deficiency Syndrome Using Flexible Random Curves},
    volume={45},
    ISSN={00359254},
    DOI={10.2307/2986151},
    number={2},
    journal={Applied Statistics},
    author={Shi, Minggao and Weiss, Robert E. and Taylor, Jeremy M. G.},
    year={1996},
    pages={151}
}

@article{Rice_Wu_2001,
    title={Nonparametric Mixed Effects Models for Unequally Sampled Noisy Curves},
    volume={57},
    rights={http://onlinelibrary.wiley.com/termsAndConditions#vor},
    ISSN={0006-341X, 1541-0420},
    url={https://academic.oup.com/biometrics/article/57/1/253-259/7276077},
    DOI={10.1111/j.0006-341X.2001.00253.x},
    number={1},
    journal={Biometrics},
    author={Rice, John A. and Wu, Colin O.},
    year={2001},
    month=mar,
    pages={253–259},
    language={en}
}

@article{James_2000,
   author={Gareth M. James and Trevor J. Hastie and Catherine A. Sugar},
   doi={10.1093/BIOMET/87.3.587},
   issn={0006-3444},
   issue={3},
   journal={Biometrika},
   month={9},
   pages={587-602},
   publisher={Oxford Academic},
   title={Principal component models for sparse functional data},
   volume={87},
   year={2000}
}

@article{James_Hastie_2001,
  title={Functional Linear Discriminant Analysis for Irregularly Sampled Curves},
  volume={63},
  ISSN={1369-7412, 1467-9868},
  url={https://academic.oup.com/jrsssb/article/63/3/533/7083370},
  DOI={10.1111/1467-9868.00297},
  number={3},
  journal={Journal of the Royal Statistical Society Series B: Statistical Methodology},
  author={James, Gareth M. and Hastie, Trevor J.},
  year={2001},
  month=sep,
  pages={533–550},
  language={en}
}

@article{chiou2003functional,
  title={Functional quasi-likelihood regression models with smooth random effects},
  author={Chiou, Jeng-Min and M{\"u}ller, Hans-Georg and Wang, Jane-Ling},
  journal={Journal of the Royal Statistical Society Series B: Statistical Methodology},
  volume={65},
  number={2},
  pages={405--423},
  year={2003},
  publisher={Oxford University Press}
}

@article{Yao_Müller_Wang_2005,
  title={Functional linear regression analysis for longitudinal data},
  volume={33},
  ISSN={0090-5364},
  url={https://projecteuclid.org/journals/annals-of-statistics/volume-33/issue-6/Functional-linear-regression-analysis-for-longitudinal-data/10.1214/009053605000000660.full},
  DOI={10.1214/009053605000000660},
  number={6},
  journal={The Annals of Statistics},
  author={Yao, Fang and Müller, Hans-Georg and Wang, Jane-Ling},
  year={2005},
  month=dec
}

@article{cardot2007conditional,
  title={Conditional functional principal components analysis},
  author={Cardot, Herv{\'e}},
  journal={Scandinavian journal of statistics},
  volume={34},
  number={2},
  pages={317--335},
  year={2007},
  publisher={Wiley Online Library}
}

@article{Paul_Peng_2011,
    title={Principal components analysis for sparsely observed correlated functional data using a kernel smoothing approach},
    volume={5},
    ISSN={1935-7524},
    url={https://projecteuclid.org/journals/electronic-journal-of-statistics/volume-5/issue-none/Principal-components-analysis-for-sparsely-observed-correlated-functional-data-using/10.1214/11-EJS662.full},
    DOI={10.1214/11-EJS662},
    number={none},
    journal={Electronic Journal of Statistics},
    author={Paul, Debashis and Peng, Jie},
    year={2011},
    month=jan
}

@article{nie2020sparse,
  title={Sparse functional principal component analysis in a new regression framework},
  author={Nie, Yunlong and Cao, Jiguo},
  journal={Computational Statistics \& Data Analysis},
  volume={152},
  pages={107016},
  year={2020},
  publisher={Elsevier}
}

@misc{Mbaka_Cao_Carey_2024,
      title={Estimation of Functional Principal Components from Sparse Functional Data}, 
      author={Uche Mbaka and Jiguo Cao and Michelle Carey},
      year={2026},
      eprint={2603.18833},
      archivePrefix={arXiv},
      primaryClass={stat.ME},
      url={https://arxiv.org/abs/2603.18833}, 
}

@article{Chiou_Yang_Chen_2014,
  title={Multivariate functional principal component analysis: A normalization approach},
  ISSN={10170405},
  url={http://www3.stat.sinica.edu.tw/statistica/J24N4/J24N45/J24N45.html},
  DOI={10.5705/ss.2013.305},
  journal={Statistica Sinica},
  author={Chiou, Jeng-Min and Yang, Ya-Fang and Chen, Yu-Ting},
  year={2014}
}

@article{Happ_Greven_2018,
  title={Multivariate Functional Principal Component Analysis for Data Observed on Different (Dimensional) Domains},
  volume={113},
  ISSN={0162-1459, 1537-274X},
  url={https://www.tandfonline.com/doi/full/10.1080/01621459.2016.1273115},
  DOI={10.1080/01621459.2016.1273115},
  number={522},
  journal={Journal of the American Statistical Association},
  author={Happ, Clara and Greven, Sonja},
  year={2018},
  month=apr,
  pages={649–659},
  language={en}
}

@article{Li_Xiao_Luo_2020,
    title={Fast covariance estimation for multivariate sparse functional data},
    volume={9},
    ISSN={2049-1573, 2049-1573},
    url={https://onlinelibrary.wiley.com/doi/10.1002/sta4.245},
    DOI={10.1002/sta4.245},
    number={1},
    journal={Stat},
    author={Li, Cai and Xiao, Luo and Luo, Sheng},
    year={2020},
    month=jan,
    pages={e245},
    language={en}
}

@article{Shi_Yang_Wang_Ma_Beg_Pei_Cao_2022,
    title={Two-Dimensional Functional Principal Component Analysis for Image Feature Extraction},
    volume={31},
    ISSN={1061-8600, 1537-2715},
    url={https://www.tandfonline.com/doi/full/10.1080/10618600.2022.2035738},
    DOI={10.1080/10618600.2022.2035738},
    number={4},
    journal={Journal of Computational and Graphical Statistics},
    author={Shi, Haolun and Yang, Yuping and Wang, Liangliang and Ma, Da and Beg, Mirza Faisal and Pei, Jian and Cao, Jiguo},
    year={2022},
    month=oct,
    pages={1127–1140},
    language={en}
}

@article{Xiao_Li_Checkley_Crainiceanu_2018,
    title={{Fast Covariance Estimation for Sparse Functional Data}},
    volume={28},
    ISSN={0960-3174, 1573-1375},
    DOI={10.1007/s11222-017-9744-8},
    number={3},
    journal={Statistics and Computing},
    author={Xiao, Luo and Li, Cai and Checkley, William and Crainiceanu, Ciprian},
    year={2018},
    month={May},
    pages={511–522},
    language={en}
}

@article{James_Sugar_2003,
    title={Clustering for Sparsely Sampled Functional Data},
    volume={98},
    ISSN={0162-1459, 1537-274X},
    url={http://www.tandfonline.com/doi/abs/10.1198/016214503000189},
    DOI={10.1198/016214503000189},
    number={462},
    journal={Journal of the American Statistical Association},
    author={James, Gareth M and Sugar, Catherine A},
    year={2003},
    month=jun,
    pages={397–408},
    language={en}
}

@article{Jacques_Preda_2014,
  title={Model-based clustering for multivariate functional data},
  volume={71},
  ISSN={01679473},
  url={https://linkinghub.elsevier.com/retrieve/pii/S0167947312004380},
  DOI={10.1016/j.csda.2012.12.004},
  journal={Computational Statistics \& Data Analysis},
  author={Jacques, Julien and Preda, Cristian},
  year={2014},
  month=mar,
  pages={92–106},
  language={en}
}

@article{Li_Chan_Doody_Quinn_Luo_2017,
    title={Prediction of Conversion to Alzheimer's Disease with Longitudinal Measures and Time-To-Event Data},
    volume={58},
    ISSN={13872877, 18758908},
    url={https://www.medra.org/servlet/aliasResolver?alias=iospress&doi=10.3233/JAD-161201},
    DOI={10.3233/JAD-161201},
    number={2},
    journal={Journal of Alzheimer's Disease},
    author={Li, Kan and Chan, Wenyaw and Doody, Rachelle S. and Quinn, Joseph and Luo, Sheng and the Alzheimer's Disease Neuroimaging Initiative},
    editor={Leoutsakos, Jeannie-Marie},
    year={2017},
    month=may,
    pages={361–371}
}

@article{Weiner_2017,
    title={The Alzheimer's Disease Neuroimaging Initiative 3: Continued innovation for clinical trial improvement},
    volume={13},
    ISSN={1552-5260, 1552-5279},
    url={https://alz-journals.onlinelibrary.wiley.com/doi/10.1016/j.jalz.2016.10.006},
    DOI={10.1016/j.jalz.2016.10.006},
    number={5},
    journal={Alzheimer's \& Dementia},
    author={Weiner, Michael W. and Veitch, Dallas P. and Aisen, Paul S. and Beckett, Laurel A. and Cairns, Nigel J. and Green, Robert C. and Harvey, Danielle and Jack, Clifford R. and Jagust, William and Morris, John C. and Petersen, Ronald C. and Salazar, Jennifer and Saykin, Andrew J. and Shaw, Leslie M. and Toga, Arthur W. and Trojanowski, John Q. and Alzheimer's Disease Neuroimaging Initiative},
    year={2017},
    month=may,
    pages={561–571},
    language={en}
}

@misc{ahasverus,
  author={Nicolas Casajus},
  title={elbow},
  year={2020},
  publisher={GitHub},
  journal={GitHub repository},
  howpublished={\url{https://github.com/ahasverus/elbow}},
  commit={cf58784f3a5c17365a1743453a3212cc07a9f216}
}

\end{document}